\documentclass[prb,aps,two column,superscriptaddress]{revtex4-1}
\usepackage{bm}
\usepackage[colorlinks=true,linkcolor=blue,citecolor=blue]{hyperref}
\usepackage{times}
\usepackage{amsmath}
\usepackage{amssymb}
\usepackage{amsthm}
\usepackage{amsfonts}
\usepackage{enumerate}
\usepackage{latexsym}
\usepackage{ifpdf}
\newcommand{\beq}{\begin{equation}}
\newcommand{\eeq}{\end{equation}}
\usepackage{graphicx}
\usepackage{makeidx}
\usepackage{color}
\usepackage{physics}
\usepackage{siunitx}

\begin{document}

\title{Oxygen vacancy induced electronic structure modification of KTaO$_3$ }
\author {Shashank Kumar Ojha}
\altaffiliation{These authors  contributed equally}
\affiliation  {Department of Physics, Indian Institute of Science, Bengaluru 560012, India}
\author {Sanat Kumar Gogoi}
\altaffiliation{These authors  contributed equally}
\affiliation  {Department of Physics, Indian Institute of Science, Bengaluru 560012, India}
\author {Prithwijit Mandal}
\affiliation  {Department of Physics, Indian Institute of Science, Bengaluru 560012, India}
\author{S. D. Kaushik}
\affiliation  {UGC-DAE-Consortium for Scientific Research Mumbai Centre, R5 Shed, Bhabha Atomic Research Centre, Mumbai 400085, India}
\author{J. W. Freeland}
\affiliation {Advanced Photon Source, Argonne National Laboratory, Argonne, Illinois 60439, USA}
\author {M. Jain}
\email{mjain@iisc.ac.in}
\affiliation  {Department of Physics, Indian Institute of Science, Bengaluru 560012, India}
\author {S. Middey}
\email{smiddey@iisc.ac.in  }
\affiliation  {Department of Physics, Indian Institute of Science, Bengaluru 560012, India}

\begin{abstract}
The observation of metallic interface between band insulators LaAlO$_3$ and SrTiO$_3$ has led to massive efforts to understand the origin of the phenomenon as well as to search for other systems hosting such two dimensional electron gases (2-DEG). However, the understanding of the origin of the 2-DEG is very often hindered as several possible mechanisms such as  polar catastrophe, cationic intermixing and oxygen vacancy (OV) etc. can be operative simultaneously.  The presence of a heavy element  makes KTaO$_3$ (KTO) based 2-DEG a potential platform to investigate spin orbit coupling driven novel electronic and magnetic phenomena.  In this work, we investigate the sole effect of the OV, which makes KTO metallic. Our detailed \textit{ab initio} calculations  not only find  partially filled conduction bands in the presence of an OV but also predict a  highly localized mid-gap state due to the linear clustering of OVs around Ta. Photoluminescence measurements indeed reveal the existence of such mid-gap state and O $K$-edge X-ray absorption spectroscopy finds electron doping in Ta $t_{2g}^*$ antibonding states.  This present work suggests that one should be cautious about the possible presence of OVs within KTO substrate in interpreting metallic behavior of KTO based 2-DEG.
  \end{abstract}

\maketitle

SrTiO$_3$-based oxide heterostructures are hosts of several emergent phenomena such as two-dimensional electron gas (2-DEG), ferromagnetism, superconductivity, etc~\cite{ohtomo:2004p423,reyren:2007p1196,mannhart:2008p1027,lee:2013p1196,brinkman:2007p493,zubko:2011p141,hwang:2012p103,lee:2013p1196,stemmer:2014p151,pai:2018p0396503}.  The larger carrier density  ($\ge$ 10$^{13}$-10$^{14}$ cm$^{-2}$ vs. 10$^{10}$- 10$^{12}$ cm$^{-2}$) and shorter confinement length (1-2 nm vs. 10 nm) of the carriers at the interface of these SrTiO$_3$ (STO) based 2-DEGs, compared to traditional semiconductor 2-DEGs makes these systems attractive for device applications~\cite{mannhart:2008p1027,chakhalian:2014p1189}. However, mobility of these STO-based heterostructures is significantly low compared to semiconductor 2-DEGs, prompting to search for non-STO based 2-DEG systems~\cite{trier:2018p293002}. Bulk KTO is a wide band gap insulator ($E_g\sim$ 3.5 eV) and has a lot of similarities with STO such as cubic structure, quantum paraelectricity~\cite{samara:1973p1256} etc.
Similar to STO, electron doping makes KTO metallic and even superconducting at very low temperature~\cite{harashima:2013p085102,ueno:2011p408,liu:2020p}. The presence of heavy element Ta  further offers possibility of achieving new emergent phases due to strong spin orbit coupling (SOC) and several experimental works about KTO based heterostructures have been reported over the last few years~\cite{nakamura:2009p121308,king:2012p117602,Shanavas:2014p086802,zou:2015p036104,thompson:2014p102901,zhang:2018p116803,zhang:2019p609,zhang:2019p1605,wadehra:2019planar,kumar:2019observation}. Comparable values of SOC strength with other energy scales in KTO leads to significant reconstruction of orbital symmetries~\cite{santander:2012p121107} at the surface which can further result non-trivial spin-orbital texturing~\cite{bruno:2019p1800860} of conduction electrons. Presence of such unconventional spin texturing of conduction electrons can be further utilized for engineering novel topological phenomena such as topological Hall effect~\cite{shashank:2020p}, making KTO based 2DEG a unique platform for topological spintronics applications~\cite{fert:2017p1}.
 
Similar to the controversial origin of the metallic behavior in  LaAlO$_3$/SrTiO$_3$ interface~\cite{nakagawa:2006p204,sing:2009p176805,herranz:2007p216803,Kalabukhov:2007p121404,park:2013p017401,von:2019p1,salluzzo:2013p2333,Breckenfeld:2013p196804}, the origin of 2-DEG in KTO based heterostructures has been linked to  electronic reconstruction due to polar catastrophe~\cite{zou:2015p036104,wadehra:2019planar}, as well as OV formation~\cite{zhang:2018p116803}. While a higher carrier density has been achieved in LaTiO$_3$/KTO interface as predicted by the polar catastrophe model~\cite{zou:2015p036104}, the presence of OVs within KTO substrate is also highly probable due to the use of high vacuum atmosphere during the growth.

In order to understand the sole effect of OV without any involvement of polar catastrophe and cationic intermixing issues~\cite{thompson:2014p102901},   we have deliberately introduced OV in KTO single crystal through thermal annealing in a reduced atmosphere. In contrast to the method of electron doping  by Ar$^+$ bombardment,  the present method does not damage crystallinity of the sample and results metallic KTO. Each OV donates two electrons into the system and our  \textit{\textit{ab initio}} calculations with isolated OV finds that both of these  electrons are doped into the  conduction band, derived from  Ta $t_{2g}*$ anti bonding states, resulting metallic behavior. We further find that  linear  clustering of OVs is more favorable than the formation of isolated OVs. Moreover, our calculations predict that such clustering would lead to  the formation of  a narrow mid-gap state between the fully filled valence band and partially filled conduction band. The existence of such mid-gap states has been confirmed by  photoluminescence (PL) measurements. Our O $K$-edge X-ray absorption spectroscopy (XAS) measurements also reveal electron doping in Ta $t_{2g}^*$ bands, as predicted by the \textit{ab initio} calculations.

\begin{figure}
	
	\includegraphics[width=.48\textwidth] {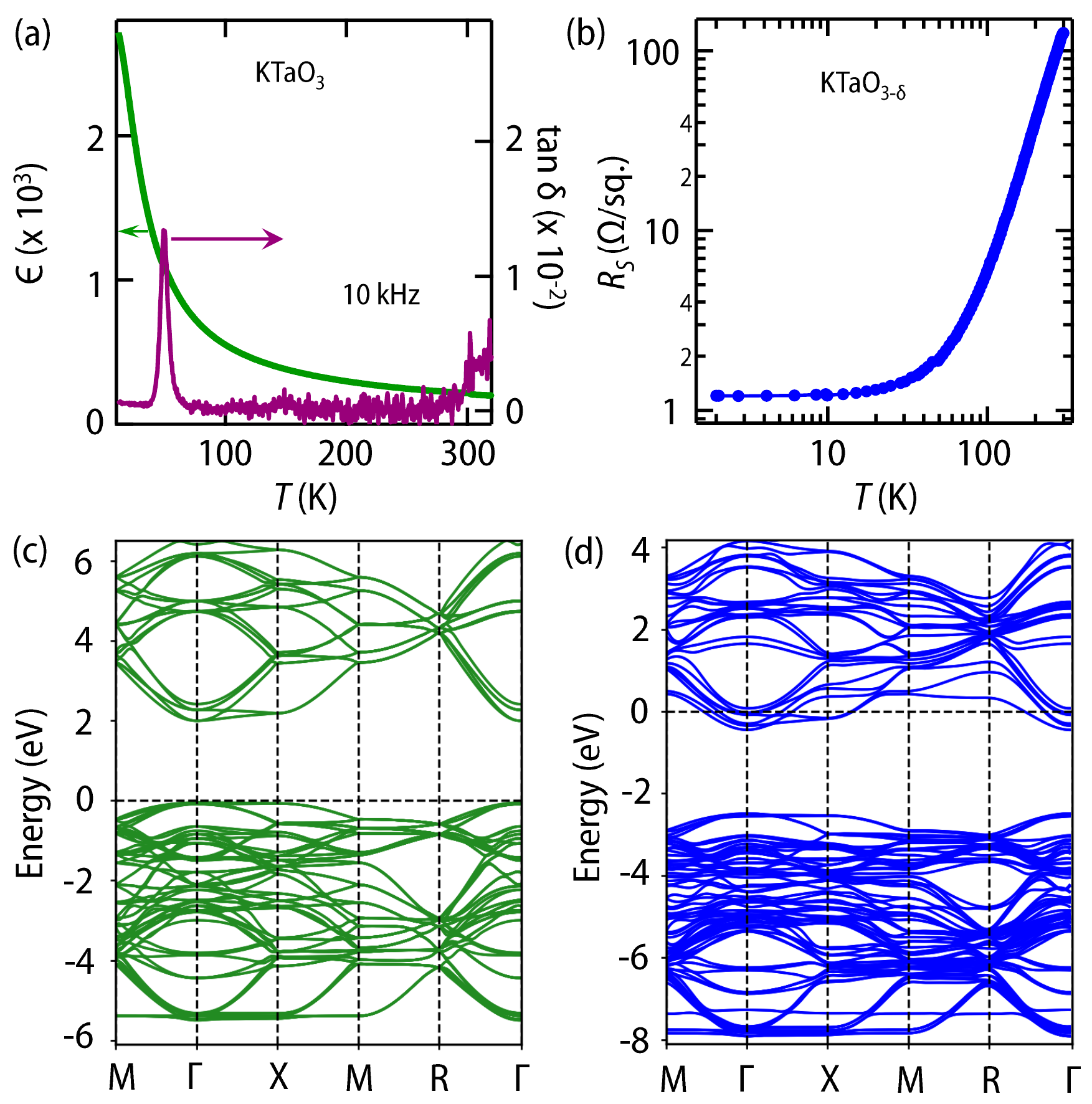}
	\caption{\label{Fig1} (Color online) (a) Temperature dependence of dielectric constant ($\epsilon$) and dielectric loss (tan $\delta$) of pristine KTO. (b) Temperature dependence of sheet resistance for oxygen deficient KTO. Band structure of (c) pristine KTO  for a 2$\times$2$\times$2 supercell and (d) oxygen deficient KTO (KTaO$_{2.875}$) for a supercell of size 2$\times$2$\times$2 with single isolated oxygen vacancy. }
\end{figure}

Commercially available (Princeton Scientific Corporation) KTO  (0 0 1) single crystal  (5 mm $\times$ 5 mm $\times$  0.5 mm) was sealed with Ti metal wire in a quartz tube under a vacuum of 10$^{-4}$ bar~\cite{middey:2012p042406}. This sealed tube was heated at 900$^\circ$C for 24 hrs. Temperature dependent dielectric measurement was carried out using impedance analyzer from Keysight technology instruments (Model No- E49908). Resistance was measured with  wire-bonded Al contact in a Van der Pauw geometry using a 9 Tesla Physical Property Measurement system (Quantum Design). XAS measurement on O $K$-edge at 300 K was carried out at 4-ID-C beamline of Advanced Photon Source, Argonne National Laboratory. PL spectra was collected in Horiba LabRAM HR instrument under excitation with a 266 nm ultraviolet laser at room temperature.
The \textit{ab initio} calculations were carried out using the QUANTUM ESPRESSO package\cite{QE-2017}. Perdew, Burke and Ernzerhof generalized gradient approximation was used for the exchange correlation functional\cite{PBE-GGA}. Optimized norm conserving pseudopotentials~\cite{Hamann:2013p085117} were used in all the calculations. The Brillouin  zone was sampled with  $8 \cross 8 \cross 8$ $k$-points for the unit cell and the wave functions were expanded in plane waves with an energy up to 90 Ry. In order to capture the effect of SOC a fully relativistic pseudopotential was used for the Ta atom. We have performed non-collinear density functional theory (DFT) calculations owing to the presence of strong SOC on Ta atom. Upon including an additional onsite Coulomb potential of 1.35 eV on the Ta $d$-orbitals, the qualitative conclusions remain unchanged \cite{UJPTa}. The structural relaxations were performed until the force on each atom reduces to 0.07 eV/\si{\angstrom}.

\begin{figure}

	\includegraphics[width=.50\textwidth] {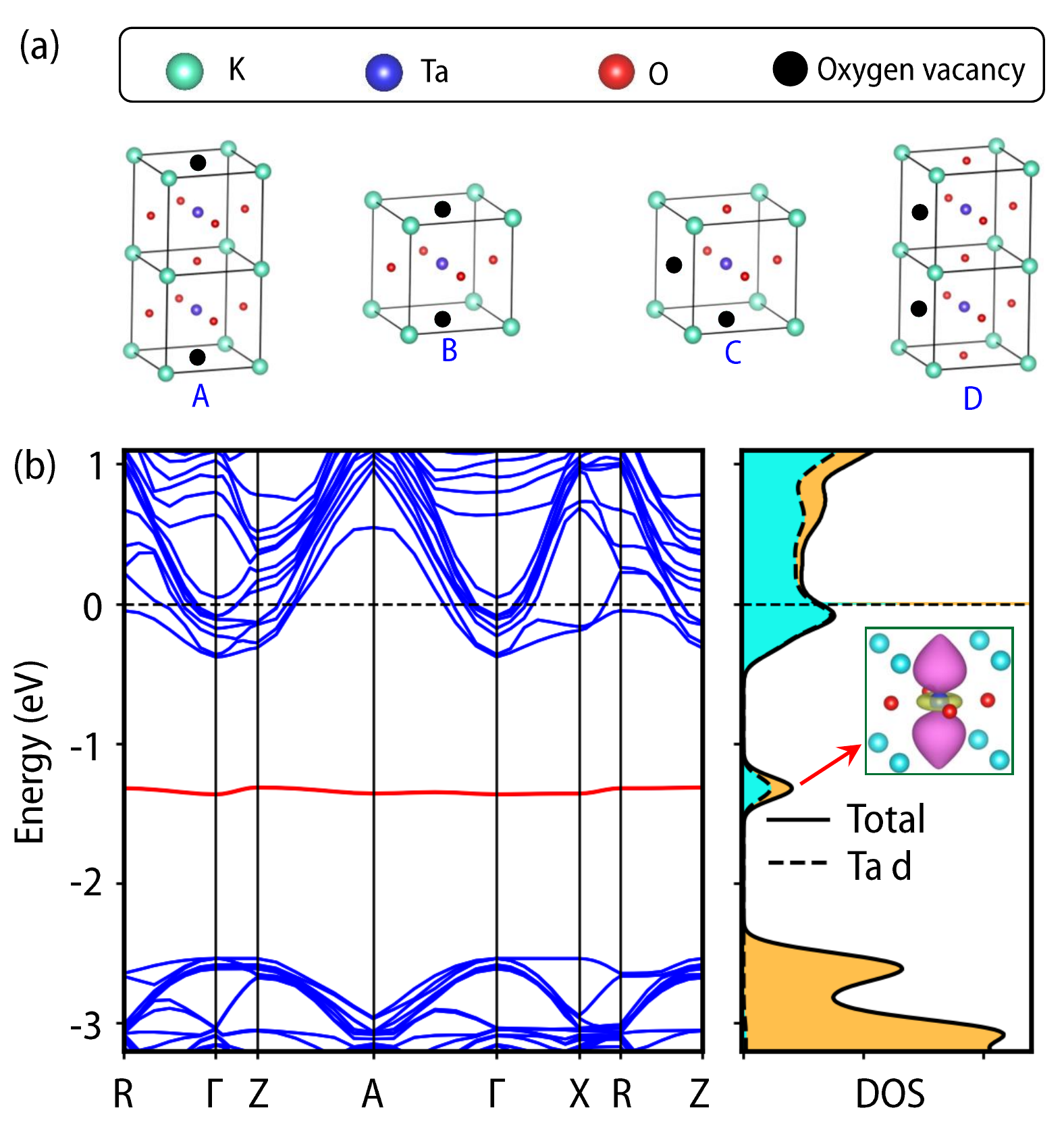}
	\caption{\label{Fig2} (Color online) (a) Various relative configurations for oxygen divacancy in KTO. (b) The band structure plot of apical divacancy (configuration B) situated along \textbf{z} direction for a $2 \cross 2 \cross 4$ supercell is shown in left panel. The defect band is marked with color red. Right panel shows the projected density of states (PDOS) for Ta 5$d$ orbitals in bluish color along with total density of states in yellow color. The inset depicts the isosurface of squired wave function for the defect band which shows Ta $d{_{3z^2 - r^2}}$ characteristics.}
\end{figure}

Fig. \ref{Fig1}(a) shows the temperature dependent dielectric constant ($\epsilon$) and dielectric loss (tan $\delta$) of  as received pristine KTO single crystal. The values of $\epsilon$ and tan $\delta$ and their temperature dependencies are very similar to earlier works~\cite{Ang:64p184104,Aktas:2014p165309}. KTO, annealed in presence of Ti wire becomes metal as evident from the temperature dependent sheet resistance ($R_S$) from 2 K to 300 K (Fig. \ref{Fig1}(b)).  Upon reheating this sample in oxygen atmosphere, we found that it becomes insulating, establishing that the origin of metallic behavior is related to the presence of OVs. The $R_S$ of oxygen deficient KTO at 300 K is about two order of magnitude smaller compared to that of 2-DEG behavior of LaTiO$_3$/KTaO$_3$~\cite{zou:2015p036104}, EuO/KTaO$_3$~\cite{zhang:2018p116803}, LaVO$_3$/KTaO$_3$~\cite{wadehra:2019planar} heterostructures. Similar to the cases of thin films grown on KTO substrates~\cite{zou:2015p036104,zhang:2018p116803,wadehra:2019planar}, $R_S$ becomes temperature independent at low temperature due to the dominant contribution of electron scattering from defects~\cite{trier:2018p293002,Verma:2014p216601}. Assuming single band transport, we found carrier density $n_S \sim$ 2$\times$10$^{15}$ cm$^2$ and  electron mobility $\mu_e\sim$ 20 cm$^2$/V-s at room temperature from Hall effect measurement (not shown).

 In order to understand the modification of electronic structure in presence of OV, we have performed electronic structure calculations. We found the lattice parameter of cubic KTO to be 4.02 \AA\, which is very close to the experimentally reported value (3.989 \AA\ ~\cite{wemple:1965pA1575}). The band structure of pristine KTO (for a 2 $\times$ 2 $\times$ 2 supercell) is shown in Fig. \ref{Fig1} (c). In this calculation, we found band gap of 2.05 eV, which is also in good agreement with previous DFT results\cite{kto:strain, kto:ktl, shing:kto}. Our calculations also found a split of the conduction bands at $\Gamma$ point by $400$ meV, as reported before~\cite{king:2012p117602,nakamura:2009p121308}. To examine the effect of OV,  we first considered isolated OV, as found  by scanning probe microscopy study on vacuum annealed KTO~\cite{Setvin:2018p572}. For the calculations with an isolated OV, we considered a $2 \cross 2 \cross 2$ supercell, which corresponds to OV concentration of 4.2\%. As evident from the  band structure plot shown in Fig.\ref{Fig1}(d), the Fermi level is shifted to the conduction band, describing the experimentally observed metallic phase of oxygen deficient KTO (density of states have been shown in Supplemental Information (SI)~\cite{sup}). Our calculations find that one isolated OV donates two electrons to the conduction band.  Among the 24 $t_{2g}$ bands (3 $t_{2g}$ bands for each Ta) in the conduction band manifold, six bands are partially filled by the two electrons.  No localized defect state is found in the gap. This conclusion remains unaffected upon increasing the supercell size to $3 \cross 3 \cross 3$.

Apart from the existence of isolated OV,  perovskite compounds like STO also show a strong tendency for  OV clustering~\cite{Muller:2004p657,Prl:main,Shanthi:1998p2153,Eom:2017p3500}. To examine such clustering, we have considered a $2 \cross 2 \cross 4$ supercell containing two OVs with 4 different arrangements of OV (see Fig.~\ref{Fig2}(a) for the relative position of two OVs). To check the stability of such defect configurations, we have calculated the defect formation energy $E_f[O_v]$, defined as~\cite{Freysoldt:2014p253}
\begin{equation}
E_f[mO_v] = E_{tot}[mO_v] + \dfrac{1}{2} m \mu_{O_2}- E_{tot}[0O_v]
\end{equation}
where $E_{tot}[mO_v]$ is the total energy of the system with $m$ number of OV, $E_{tot}[0O_v]$ is the total energy of the pristine cell and $\mu_{O_2}$ is the chemical potential of oxygen molecule. Formation energies for isolated OV as well as oxygen divacancy for different configurations are listed in the TABLE \ref{tab:table1}. The oxygen defect formation energies are in the same range that have been reported for other perovskite oxides \cite{Astala_2001,EGLITIS:20161}. Moreover, we found that the linear oxygen divacancy around a Ta atom (configuration B) has the lowest formation energy which suggests that in a real system the formation of a linear oxygen divacancy has a higher probability than two isolated vacancies.

\begin{table}[h]
	
	\caption{Table of formation energy $E_f$ per OV, and interaction energy $E_{int}$ for different oxygen divacancy configurations in KTO.}
		\label{tab:table1}
		\begin{tabular}{c c c}
			\hline
			\text{Configuration} & \text{Formation Energy} & \text{Interaction Energy} \\
			&  $E_{f}$ (eV) & $E_{int}$ (eV) \\
			\hline
			A    &  6.6  &   0.9  \\
			B    &  5.9  &
			
			-0.5  \\
			C    &  6.4  &   0.4  \\
			D    &  6.7  &   1.0  \\
			single OV    &   6.8&  -- \\
			\hline
			\hline
		\end{tabular}		
	
\end{table}

Owing to the fact that the divacancies have lower formation energy compared to two isolated OVs, we also calculated the interaction energy $E_{int}$ between the vacancies which is defined as~\cite{Prl:main}
\begin{equation}
E_{int} = E_{tot}[2O_v] + E_{tot}[0O_v] - 2E_{tot}[1O_v]
\end{equation}
The negative (positive) sign of $E_{int}$ specifies whether the interaction is attractive (repulsive). The interaction energies for different divacancy configurations are listed in the TABLE \ref{tab:table1}. As can be seen,  the  configuration B,  also  known as apical divacancy (OV-Ta-OV)  has the most attractive interaction, implying that it is the  most favorable configuration among all divacancy configurations. Our calculations demonstrate that the linear vacancy clustering is  favorable around Ta. It should be noted that even though the configurations B and D have the same separation between the two vacancies, the $E_{int}$ is very different. This implies that the location of the vacancies is important and not just the relative distance.

\begin{figure}
	\hspace{-0.38cm}
	\includegraphics[width=.48\textwidth] {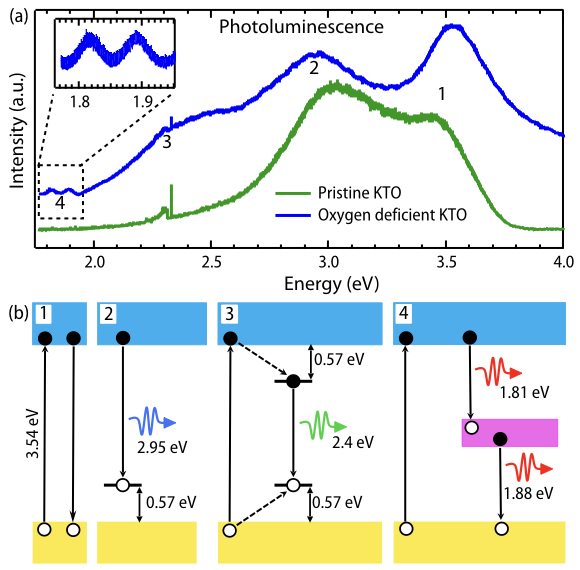}
	\caption{\label{Fig3} (Color online) (a)  PL spectra of pristine and oxygen deficient KTO. The inset shows the zoomed in view of mid-gap states around 1.8 eV. (b) The first panel corresponds to the near band edge emission. The second panel shows the recombination process between the excited electrons and the self-trapped excited holes. The third panel demonstrates the recombination process of excited holes and electrons in a self-trapped excited state. The fourth panel highlights the recombination process via defect band resulting in the peak around 1.8 eV in PL. Filled circle denotes electrons and open circle denotes holes.}
\end{figure}

Left panel of Fig. \ref{Fig2}(b) shows the band structure plot of configuration B and  projected density of states (PDOS) right next to it with the same energy scale used in the band structure plot.  Apart from the partially filled conduction bands, we find a highly localized defect band (marked in red color), which is almost equidistant from conduction and valence band edges. This localized defect state is occupied. The PDOS plot shows that the defect band is mainly contributed by the Ta 5$d$ states. The isosurface plot of the squared wave function  of the defect band  (inset of Fig. \ref{Fig2}(b))  is localized around the Ta atom in between the two OVs.  The $d_{3z^2 - r^2}$ symmetry of this occupied defect band is related to the fact that the  particular Ta atom with oxygen divacancy has square planar environment locally and the $d_{3z^2 - r^2}$  is the lowest orbital for square planar crystal field~\cite{MorettiSala:2011p043026}. Highly localized defect band has been also obtained in band structure calculation for the configurations A and C (shown in SI~\cite{sup}). We have also verified our results with a bigger supercell of size $4 \cross 4 \cross 4$ and the overall results remain unchanged.

In order to check the presence of such mid-gap state,  we did  PL measurement as this technique has been found to be very successful in locating the exact position of such defect states in oxygen deficient STO~\cite{Kan:2005p816,Crespillo:2017p155303,Xu:2013p154106,jay:2011p035101,jay:2010p3983}. Fig. \ref{Fig3}(a) shows the room-temperature PL spectra for the pristine and oxygen deficient KTO. The corresponding mechanism for the peaks marked by  1, 2, 3, and 4 have been shown in Fig. \ref{Fig3}(b).    The peak at 3.5 eV for the pristine sample corresponds to the near band edge emission. Position of this peak directly corresponds to the band gap of KTO and matches very well with the reported band gap from UV-visible spectroscopy\cite{wemple:1965pA1575}. Since, this transition occurs from bottom of conduction band to top of valence band, this peak is also observed for oxygen deficient KTO. KTO is an incipient ferroelectric~\cite{samara:1973p1256} and has strong electron-phonon coupling~\cite{Katayama:2006p064713}.  In presence of electron-phonon coupling, photo generated electron and hole pairs are quickly trapped by phonons to form more stable self-trapped electron/hole states~\cite{Li:2019p1999,Menzel:1990p163,Scholz:2005p245208,Scholz:2005p245208}. Recombination between these self-trapped electron and hole centres  ( $3^{\text{rd}}$ panel of \ref{Fig3}(b)) leads to the  `green' luminescence around 2.4 eV, which has  been  reported extensively in various perovskite oxides~\cite{eglitis:2003} including KTO~\cite{Katayama:2006p064713}. This feature is strongly enhanced in present case upon introduction of OV, as reported earlier for STO~\cite{mochizuki:2005p923}. The peak at 2.95 eV can be attributed to the recombination of excited and conduction electrons with the self-trapped holes ($2^{\text{nd}}$ panel of the Fig. \ref{Fig3} (b)).

\begin{figure}
	\hspace{-0.3cm}
	\includegraphics[width=.50\textwidth] {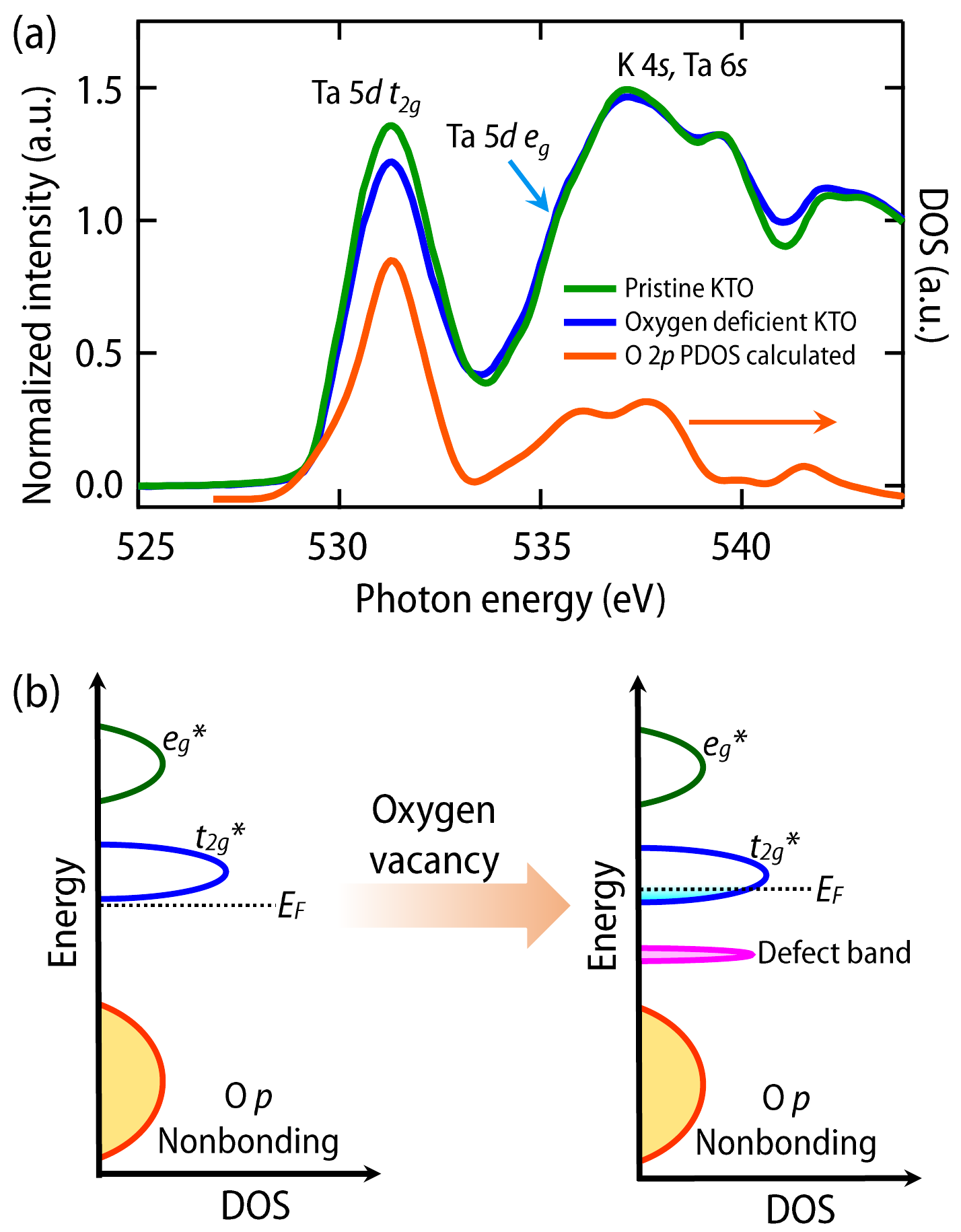}
	\caption{\label{Fig4} (Color online) (a) O $K$-edge XAS spectra for pristine and oxygen deficient KTO along with O 2$p$ PDOS for pristine KTO. (b) A schematic to show
		electronic structure modification of KTO due to oxygen vacancies.}
\end{figure}

Apart from these  broad features, two new  peaks have been observed at 1.81 eV and 1.88 eV (also see inset of Fig. \ref{Fig3}(a)) upon the creation of OV. The origin of these peaks can be understood by considering a defect band, which would be at 1.8 eV below the minima of conduction band ($4^{\text{th}}$ panel of Fig. \ref{Fig3}(b)).  This mid-gap state is separated equally from the valence and conduction band. As obtained from calculations, such localized mid-gap states can only be accounted by considering clustering of OVs. Furthermore, the band structure plot (Fig. \ref{Fig2}(b)) for the most favorable apical divacancy case (configuration B in Fig. \ref{Fig2}(a)) has also demonstrated that  the defect state is in the middle of the valence band and conduction band. This strongly implies  the presence of linear clustering of vacancies in our oxygen vacant KTO sample.

PL measurements provides information of energy levels between the top of valence band and the bottom of conduction band. In order to understand the effect of oxygen vacancy in the unoccupied density of states, we measured O $K$-edge XAS, where one core electron from O 1$s$ state is excited to O 2$p$ states~\cite{Suntivich:2014p1856}. In an ionic picture, such O 1$s\rightarrow$ O 2$p$ transition is not allowed as 2$p$ orbitals are completely occupied for the O$^{2-}$ ion. However, the strong hybridization between oxygen and other elements in a real material gives rise to a finite spectral weight of oxygen 2$p$ character in the unoccupied density of states, which can be approximately described by O $K$-edge XAS~\cite{sarma:1996p1622}.  Normalized XAS spectra, recorded in bulk sensitive total fluorescence yield (TFY) mode have been compared in Fig. \ref{Fig4}(a). To understand the origin of various features of these XAS spectra, convoluted density of states  of O 2$p$ states has been also plotted (for details see SI~\cite{sup}).  The first peak around 531.3 eV is due to transition to the states just above Fermi level and primarily consists of Ta 5$d$ $t_{2g}$  orbital, hybridized with O 2$p$ orbitals~\cite{Kuepper:2004p8213}.   The lower intensity of this peak in oxygen deficient sample, compared to the pristine KTO implies that some of the doped electrons have occupied Ta $t_{2g}*$ states, which is also concluded by our calculations shown in Fig. \ref{Fig1} and \ref{Fig2}. The features from 534 eV to 544 ev are related to the transitions to the   Ta 5$d$ $e_{g}$, Ta  6$s$, K 4$s$ states hybridized with O 2$p$ (see SM). Finally, combining the results of DFT calculations, PL and O $K$ edge XAS measurements, we show a schematic (Fig. \ref{Fig4}(b)) to summarize the electronic structure modification of KTO due to OVs.

To conclude, we have successfully demonstrated the sole effect of oxygen vacancy on electronic structure  of KTO. Pristine insulating KTO undergoes insulator to metal transition (IMT) upon oxygen vacancy creation. DFT calculations shows that IMT can be explained by just considering isolated oxygen vacancies. Further our DFT calculations combined with photoluminescence measurements reveal that linear clustering of oxygen vacancies around Ta atom leads to the formation of very localized state within the band gap, which can result fascinating magnetotransport phenomena in KTO~\cite{Lopez:2015p100701}.

SKO and SM thank Professor D. D. Sarma for giving access to several experimental facilities for this work. We acknowledge Sayak Mondal, Ashutosh Mohanty,  Subhadip Das and Dr. D. V. S. Muthu for help with the experiments. SKG and MJ acknowledge Dr. Tathagata Biswas for fruitful discussions. This work is funded by a DST Nanomission grant (DST/NM/NS/2018/246), and a SERB Early Career Research Award (ECR/2018/001512). SM also acknowledges support from Infosys Foundation, Bangalore. The authors are grateful to Supercomputer Education and Research Centre, IISc for providing computational facilities. This research used resources of the Advanced Photon Source, a U.S. Department of Energy Office of Science User Facility operated by Argonne National Laboratory under Contract No. DE-AC02-06CH11357.

\clearpage
\setcounter{figure}{0}

\renewcommand{\thefigure}{S\arabic{figure}}
\onecolumngrid 
\makebox[\textwidth]{\bf \Large Supplemental Information}

\hspace{1cm}

\makebox[\textwidth]{\bf \Large Oxygen vacancy induced electronic structure modification of KTaO$_3$}
\hspace{1cm}

\makebox[\textwidth]{\large Shashank Kumar Ojha, Sanat Kumar Gogoi, Prithwijit Mandal, S. D. Kaushik, J. W. Freeland, M. Jain,  and S. Middey}

\hspace{1cm}

{\bf \large Density of state (DOS) for pristine KTO and KTO with single oxygen vacancy:} Fig S1 (a) shows the total DOS for the pristine KTO along with PDOS for Ta 5$d$ (yellow in color) state and O 2$p$ (green in color) state. The states below -2 eV are Ta $d$-O $p$ bonding states and O $p$ non bonding states are located within -2 eV and 0 eV. Fig S1 (b) shows the DOS for KTO with single oxygen vacancy of supercell size  $2 \times 2 \times 2$. The conduction band manifold is contributed by Ta 5$d$ antibonding states.

\begin{figure} [h]
	\vspace{-0pt}
	\includegraphics[width=0.9\textwidth] {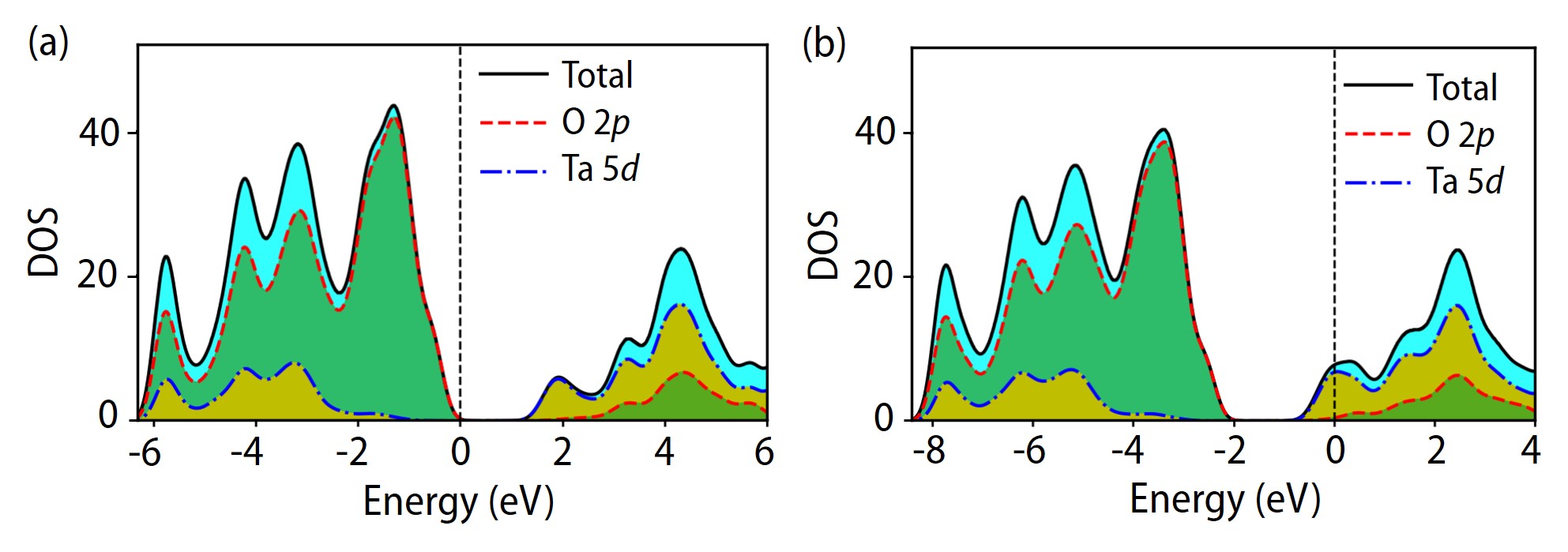}
	\caption{\label{FigS1}(a) DOS for the pristine KTO. (b) DOS for KTO with  single oxygen vacancy in the supercell of size $2 \times 2 \times 2$. }
\end{figure}

% \hspace{1cm}

\newpage
{\bf \large Band structure plot for configuration \textbf{A} and \textbf{C}:} Fig. S2 (a) shows the band structure for the configuration \textbf{A} (also see Fig. 2 of main text for the definition of various divacancy configurations). We have found the mid-gap state (shown in color red) for this configuration of oxygen vacancy also, however it should be noted that the position of the defect state is shifted towards the conduction band manifold compared to the lowest energy configuration \textbf{B}. We have also found mid-gap state for configuration \textbf{C} as shown in Fig. S2 (b).  If we compare the gap $\Delta E$  between the mid-gap state and the conduction band minima then it follows the trend: $\Delta E_C < \Delta E_A < \Delta E_B$.

\begin{figure}[h]
	\vspace{-0pt}
	\includegraphics[width=0.9\textwidth] {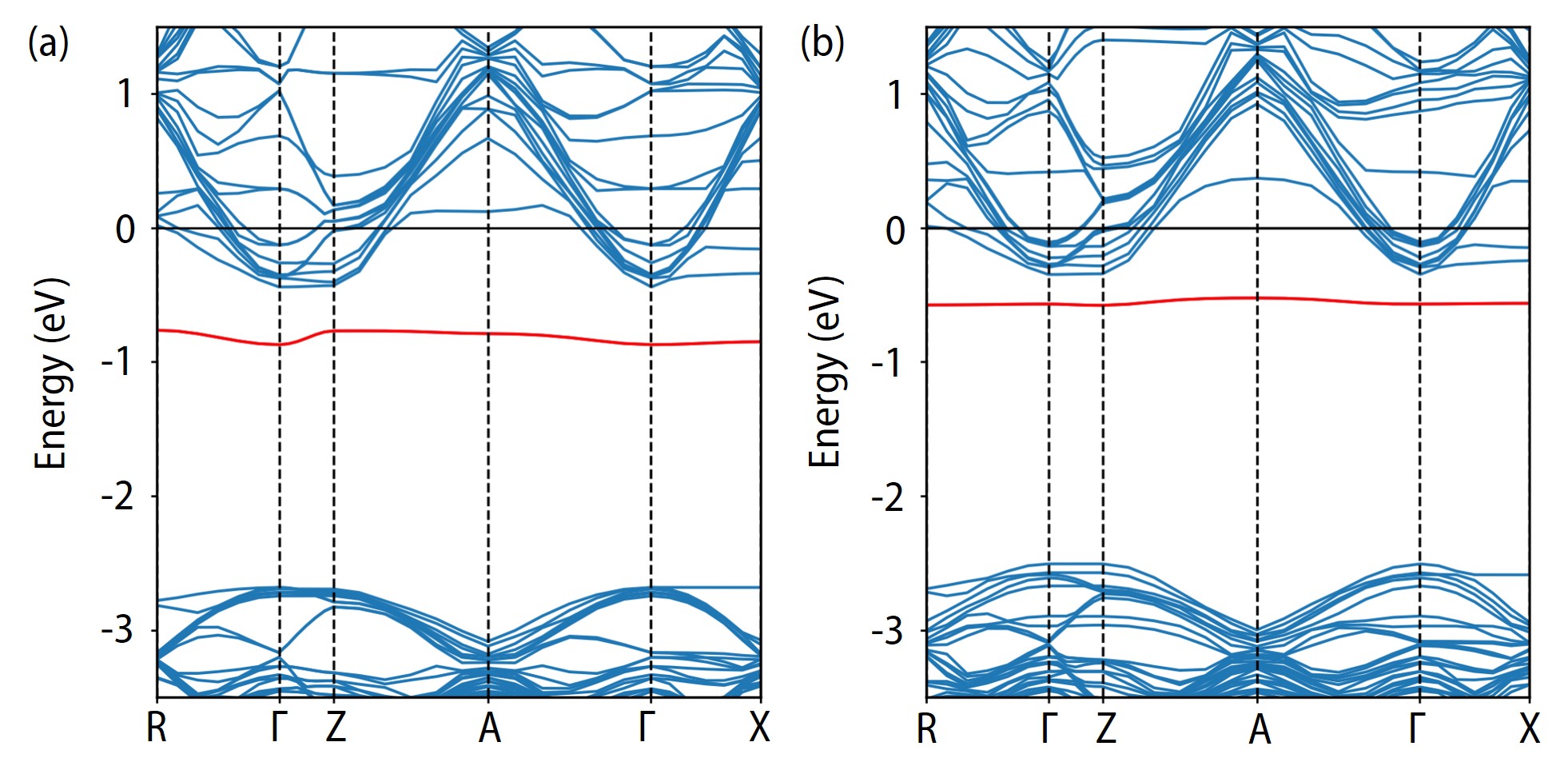}
	\caption{\label{FigS2}(a) Band structure for configuration \textbf{A}. (b) Band structure for configuration \textbf{C} }
\end{figure}

{\bf \large Orbital characters of bands above Fermi level:}
From the density of states plot (Fig. S3), it is evident that states just above the Fermi level are Ta 5$d$ $t_{2g}$ states. The states with 6.5 eV to 17 eV are contributed by Ta 5$d$ $e_{g}$, Ta 6$s$ and K 4$s$. All of these states are strongly hybridized with O $2p$ states.
\begin{figure} [h]
	\vspace{-0pt}
	\includegraphics[width=0.6\textwidth] {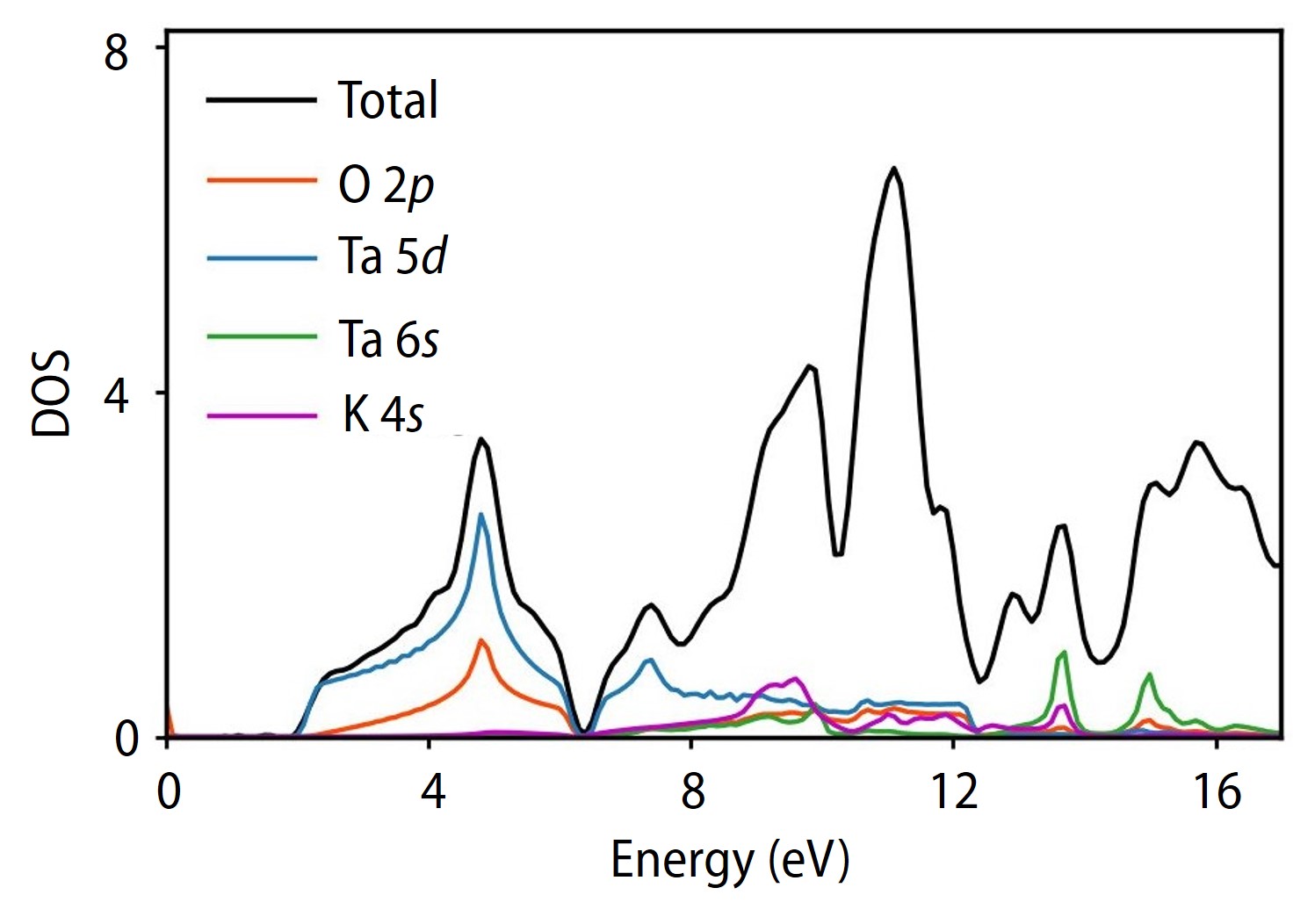}
	\caption{\label{FigS3} Total DOS along with PDOS for O 2$p$, Ta 5$d$, Ta 6$s$ and K 4$s$ for KTO unit cell. Zero of the energy refers to Fermi level.}
\end{figure}

 \end{document}